# A Multiobjective Tabu Framework for the Optimization and Evaluation of Wireless Systems

Katia Jaffrès-Runser[1,2], Jean-Marie Gorce[1] and Cristina Comaniciu[2]
*[1] CITI Laboratory, INSA Lyon*
*[2] Stevens Institute of Technology, Hoboken, NJ*
*[1]France*
*[2]United States of America*

## 1. Introduction

This chapter provides an insight on the use of Tabu search for the resolution of multiobjective (MO) optimization problems. Many real-world engineering problems are not defined with a single objective. Their complexity and the user demands lead to the use of multiple performance criteria that have to be optimized concurrently. For instance, when optimizing the shape of aerodynamic systems (Jaeggi et al., 2008) or when looking for the optimal placement of base stations in cellular networks (Reininger & Caminada, 2001), several optimization functions are necessary to account for the different performance metrics or services rendered by the system. In contrast to mono-objective optimization problems, there is rarely a unique solution that optimizes all the criteria. This is due to the fact that the objectives are often in conflict and that several feasible trade-off solutions exist, representing the best available feasible configurations of the system. Depending on the goal of the designer, one or several of these trade-off solutions can be eventually applied to the real implementation. This set of trade-off solutions is known as the Pareto-optimal set or as the optimal Pareto front.

The role of a MO optimization algorithm is to find the best possible representation of the Pareto front. Due to the complexity of the evaluation functions in real-world systems, and due to the high number of continuous or discrete variables that usually characterizes these optimizations, an exact resolution of the problem is rarely affordable within a reasonable computation time. Therefore, most of the search algorithms are based on well-known metaheuristics such as genetic algorithms, simulated annealing or local search techniques.

Main efforts in developing MO algorithms have been devoted to adapt genetic algorithms and evolution strategies to multiobjective optimization. The first multiobjective genetic algorithm was developed in 1985 and research in this field has been very active since (Deb, 2001). Simulated annealing techniques have also often been proposed. However, few MO search metaheuristics rely on Tabu search. A 2002 survey on multiobjective optimization techniques (Jones et al., 2002) showed at that time that Tabu search inspired techniques only represented 6% of the literature investigated (the survey was conducted over 115 articles), while 70% of the articles proposed a genetic / evolution-based strategy and 24% proposed a simulated annealing implementation. However, Tabu Search (TS) gained more attention in the last years as several seminal works on MO Tabu Search have been published.

The purpose of this chapter is to provide the reader with a clear view on the use of TS in the field of multiobjective optimization and to outline its performance for real-world optimization problems. After a review of the main Tabu inspired MO optimization algorithms, a simple MO Tabu Search procedure developed by the authors (referred to as PMOTS) is introduced. This heuristic has been developed to resolve the wireless LAN access point planning problem (WLP problem). The WLP planning problem is used throughout the whole chapter as an illustrative example for the statements presented herein. Another real-world optimization problem in the field of wireless networks is also presented at the end of the chapter to provide another implementation example. This new example deals with the problem of performance evaluation for routing in a wireless sensor network, where routing performance benchmarks can be provided by the search for the Pareto optimal data forwarding patterns in the network.

The outline of this chapter is the following: Section 2 introduces the WLP problem as a multiobjective optimization problem. Section 3 presents the main concepts of multiobjective optimization and gives an overview of the main MO strategies developed so far. Section 4 concentrates on the use of Tabu Search for multiobjective optimization. It first presents the Tabu based MO metaheuristics found in the literature and then focuses on the description of the proposed MO-Tabu heuristic. In section 5, a discussion on the adaptation of MO-Tabu for the WLP resolution is given and the related results are presented in a first subsection. Then, the use of PMOTS for the evaluation of wireless sensor networks is discussed. Section 6 provides concluding remarks concerning the use of Tabu in a multiobjective optimization context.

## 2. MO optimization for wireless systems

### 2.1 The Wireless LAN Planning problem (WLP problem)

In the last decade, wireless LANs have experienced great success as lots of networks have been deployed in companies or private areas either as hot spots for public access or as private networks. More and more mobility-related applications such as Voice over IP for WiFi networks are being implemented on WLANs. The locations of the access points (AP) in the service area are key factors for design and strongly influence the performance of the network. For small networks, simple rules of thumbs (e.g. site surveys, quick installations, on-site network tuning) can be applied to plan the network. However, when the network grows (i.e. more than 10 APs) and a higher quality of service is required at the application level, automatic planning tools are needed to tackle such a problem.

In WLAN systems, bad performance originates from three main reasons: a lack of radio coverage in the design area (i.e. low Signal to Noise Ratio (SNR)), inter-cell interference (i.e. low Signal to Interference and Noise Ratio (SINR)) and a high number of users sharing the same AP. WLAN planning thus aims at coping with all of these issues by looking for the network configuration that minimizes one or more evaluation criteria. In the literature, early works mostly dealt with coverage constraints (Sherali et al., 1996). Interference being a strong limiting factor for wireless networks, interference mitigation constraints has then been introduced in the problem formulation (Aguado-Agelet, 2002). Finally, throughput considerations are also accounted for in the problem definition to ensure a substantial quality of service in the network (Bahri & Chamberland, 2005).

The WLP problem is clearly a multiobjective optimization problem. Most of the times, it has been formulated as a discrete minimization problem since the set of available locations for the APs is finite as it depends on geometry of the deployment area.

## 2.2 Formal definition of the WLP problem

The following combinatorial WLP problem definition has been proposed by the authors in (Jaffrès-Runser et al., 2008). A simplified description of the problem is provided herein. See (Jaffrès-Runser et al., 2008) for a more detailed description.

Let us define a discrete set of $M$ candidate AP locations in a building floor (cf. Figure 1). Variables of the planning problem are the number $N$ of APs to plan, their locations, their transmission powers and their directions of emission if the antennas are directional. The number of APs to plan, $N$, can either be fixed or variable. In the latter case, it can be minimized to reduce the deployment costs and become an objective of the problem. However, in our model, we do not explicitly define the minimization of $N$ as an objective and rather use the concurrent optimization of coverage and interference to obtain a value of $N$ that provides enough coverage of the building without inducing too much interference.

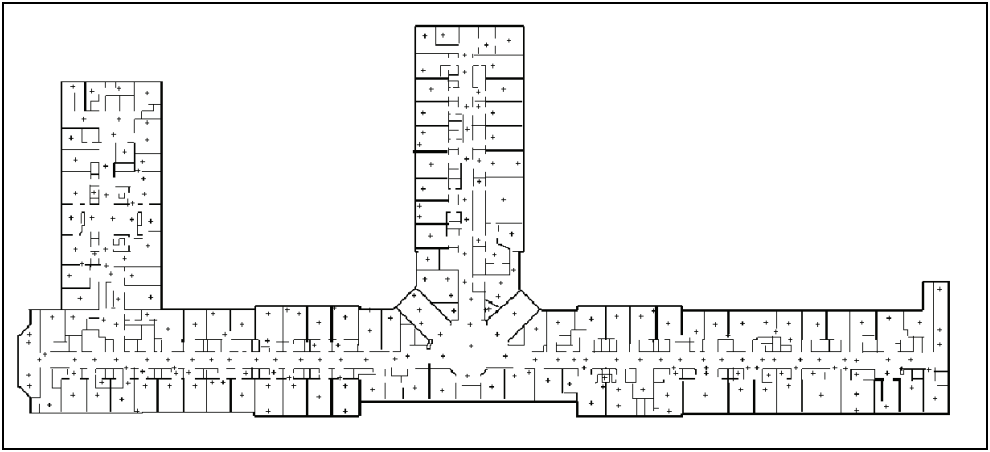

Fig. 1. Distribution of the $M = 256$ candidate AP locations on a 12400 m² building floor.

A set of $M$ candidate AP locations is defined over the building floor as represented by the crosses in Figure 1. A solution $S$ of the optimization problem is defined as a vector of $M$ items $S = (s_1, \ldots, s_i, \ldots, s_M)$, each item $s_i$ representing a candidate AP located at point $i$ in the building. If the candidate AP at location $i$ is selected in the solution $S$, $s_i$ stores the transmission power $p_i$ and direction of emission $d_i$. The item $s_i$ is defined by:

$$s_i = \begin{cases} (p_i, d_i) & \text{if an AP is placed at the } i\text{th candidate location,} \\ 0 & \text{otherwise.} \end{cases} \qquad (1)$$

Transmission power values $p_i \in \mathcal{P} = \{P_1, \ldots, P_{NP}\}$ and directions of emission to $d_i \in \mathcal{D} = \{D_1, \ldots, D_{ND}\}$ belong to two discrete sets of values. Given a directive antenna, each $D_i$ represents a possible orientation of the main lobe of the radiation pattern. In the results presented in Section 5, an omni-directional antenna is used and thus $d_i$ is fixed.

For each candidate location, a 2D coverage map is computed with a specific propagation simulator (De La Roche et al, 2007). A coverage map is defined as a set of mean received signal powers associated with rectangular areas defined by their top left-hand corner $p=(x,y)$ and dimension $s=(s_x, s_y)$ in pixels. For the sake of clarity, these areas are numbered from *1* to *L*, and referred to as the blocks $B_l$. The mean received signal power in *dBm* (logarithmic scale) from an AP numbered *k* on the block $B_l$ is referred to as $F_l^k$.

Three optimization criteria are defined in (Jaffrès-Runser et al., 2008). The first one ensures that coverage is provided for all the blocks $B_l$ of the design area. A block $B_l$ is considered covered if $F_l^k$ meets a given threshold. The second criterion ensures that the power of interference due to the APs that are not serving block $B_l$ is limited. And the third quality of service (QoS) criterion ensures that the throughput available on block $B_l$ is above a fixed threshold value.

For each criterion (coverage, interference mitigation or throughput), a specific utility value $U_l$ is assigned to a block $B_l$. This utility value is derived from the power values $F_l^k$ and depends on the kind of evaluation performed. For the coverage criterion, this value represents the maximum received power value on the block $B_l$. For the interference criterion, this utility value is the power of the second best AP received in the block (it is the strongest interfering signal). For the QoS criterion, the utility is the estimated throughput $d_l$ obtained in that block $B_l$ assuming a uniform distribution of $N_r$ users in the building.

With such a definition, a good solution in terms of coverage is a solution where there is no lack of coverage. A good solution in terms of interference is a solution where the power of the signals that interfere with the main AP is minimized. A good solution in terms of QoS is a solution where the throughput provided to the user meets a minimum threshold value.

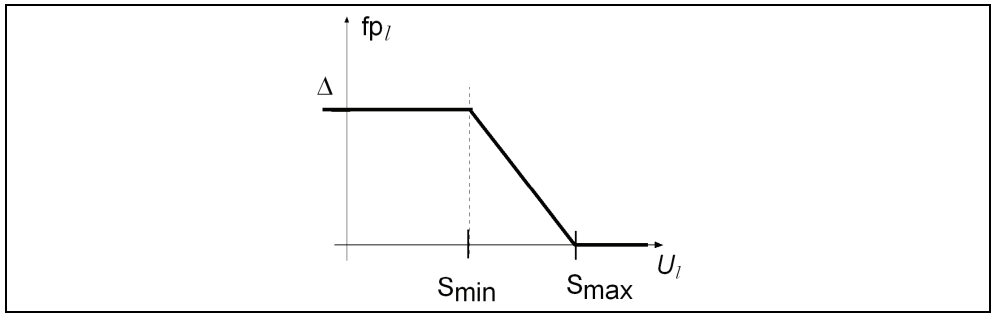

Fig. 2. Value of the penalty $fp_l$ when the utility value $U_l$ is minimized.

A penalty value $fp_l$ estimating the quality of $B_l$ regarding a specific criterion is computed. This value penalizes a block whose utility value $U_l$ does not meet the constraints for that particular criterion (*cf.* (Jaffrès-Runser et al., 2008) for further descriptions). This penalty is a function of the utility value $U_l$ on the block $B_l$ and is depicted on Fig.2. When minimizing utility $U_l$, we have a penalty equal to 0 when $U_l$ is higher than a threshold $S_{max}$. In this case, the constraint on $U_l$ is fully met. A maximum penalty of $\Delta$ is applied when the utility becomes lower than the $S_{min}$ threshold. A linear penalty is considered in between.

Each optimization criterion is defined as a quadratic weighted sum of the penalty values accounted for each block $B_l$:

$$f = \sqrt{\sum_{B_l, l \in \{1, \ldots, L\}} \mu_l . \mathtt{f} \mathtt{p}_l^2}. \qquad (2)$$

This quadratic representation allows us to concurrently minimize the average penalty together with the standard deviation values for the penalties, thus ensuring a more homogeneous distribution of coverage, interference and throughput. The notations $f_{cov}$, $f_I$ and $f_{QoS}$ are used in the following to address the coverage, interference mitigation and QoS provision criteria, respectively.

Each objective measures the quality of one feature of the network. Getting the solution that has the best possible rating for each criterion is barely possible, especially as the optimization criteria have an antagonistic influence on the solutions. For instance, networks made up of a high number of APs have good coverage and throughput performance but suffer from high interference levels.

## 3. Multiobjective optimization

### 3.1 Definitions

This section summarizes the main concepts of multiobjective optimization. First, we introduce the dominance relation between two solutions $\mathbf{x} \in S$ and $\mathbf{y} \in S$ of an $n$-objective MO problem in definition 1. Then, definition 2 defines the Pareto optimality and definition 3 the optimal Pareto front of a MO problem. The difference between the optimal Pareto front and the estimated Pareto front is presented in definition 4 and finally, definition 5 explains the notion of Pareto rank.

**Definition 1:** *A solution $\mathbf{x}$ dominates a solution $\mathbf{y}$ for a n-objective MO problem if $\mathbf{x}$ is at least as good as $\mathbf{y}$ for all the objectives and $\mathbf{x}$ is strictly better than $\mathbf{y}$ for at least one objective.* Mathematically, we have for a minimization problem:

$$\forall i \in [1, n] : f_i(\mathbf{x}) \leqslant f_i(\mathbf{y}), \quad \exists j \in [1, n] : f_j(\mathbf{x}) < f_j(\mathbf{y}) \qquad (3)$$

**Definition 2:** *A solution $\mathbf{x} \in S$ is Pareto optimal if there is no other solution $\mathbf{y} \in S$ that dominates $\mathbf{x}$.*
A Pareto-optimal solution is therefore a non-dominated solution of the problem.

**Definition 3**: *The optimal Pareto front $\mathcal{F}^*$ of a multiobjective problem is defined as the set of Pareto-optimal solutions (or the set of non-dominated solutions).*

The task of a multiobjective optimization search is to find the Pareto optimal front $\mathcal{F}^*$. The set of non-dominated solutions found at the end of the search represents the best approximation known for the Pareto optimal front. It is defined as follows:

**Definition 4:** *An estimated Pareto front $\mathcal{F}_P$ of a multiobjective problem is the set of non-dominated solutions obtained at the end of a search performed by any heuristic. The search succeeds if the estimated Pareto front is equal to the optimal Pareto front, i.e. $\mathcal{F}_P \equiv \mathcal{F}^*$.*

Figure 3 represents the optimal Pareto front obtained for a multiobjective minimization problem of two criteria. This figure is obtained after an exhaustive search of all the solutions of an instance of the WLP problem. As the search is exhaustive, the estimated Pareto front is equal to the optimal Pareto front.

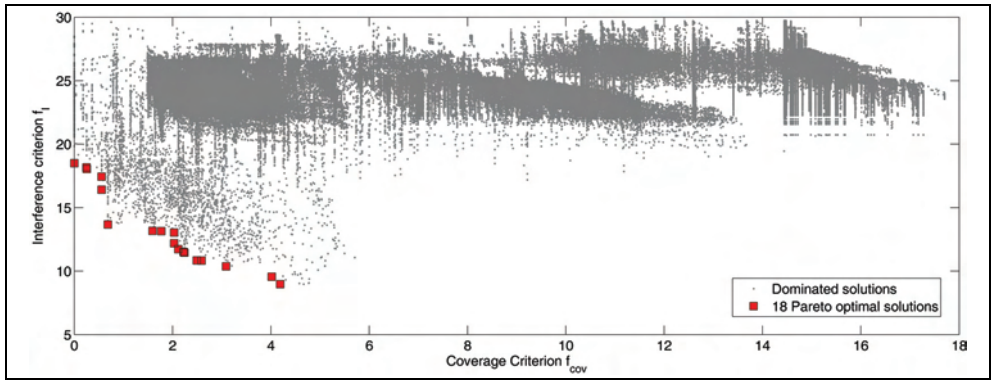

Fig. 3. Illustration of the Pareto set for a 2-objective minimization problem: the WLP problem with coverage and interference mitigation criteria.

In this problem instance, the coverage and interference mitigation criteria are minimized for a small deployment environment of 2100m². There are $M$=129 candidate AP sites and the number of APs to plan is fixed to $N$=3. Each point on this graph represents the evaluation of one solution for both criteria. The black points are the solutions dominated by the solution of the Pareto front. The desirable solutions are all the solutions depicted with red squares which provide the best available trade-offs between coverage and interference mitigation. The solutions of an MO problem can be sorted according to their Pareto rank defined as follows:

**Definition 5**: *The rank of a solution* **x** *is defined by* $R(\mathbf{x}) = 1 + d(\mathbf{x})$, *where* $d(\mathbf{x})$ *is the number of solutions by which* **x** *is dominated in the set of feasible solutions S. The solutions of the theoretical Pareto front have a rank $R(\mathbf{x}) = 1$.*

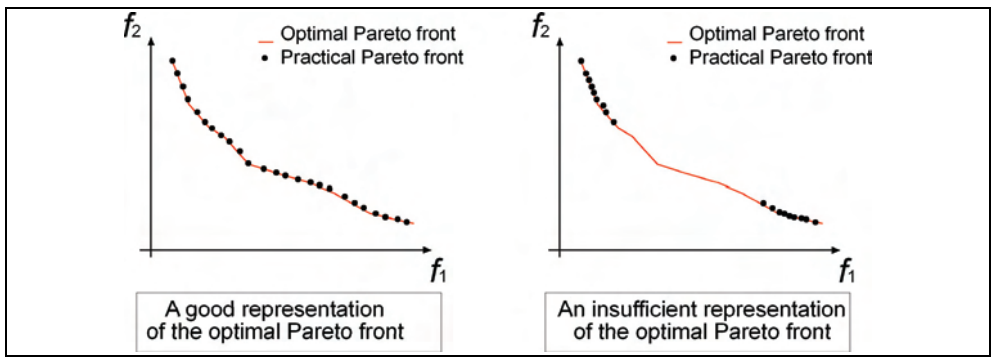

Fig. 4. Good and insufficient representation of the optimal Pareto front.

It is not always possible and necessary to search for all the solutions of the Pareto front. However, the designer needs a good representation of the Pareto front to take its decision. Therefore, heuristics have to provide an estimated Pareto front that displays the main trade-offs of the optimal front. To that end, the estimated front must be made of a set of solutions that evenly spans the whole front as depicted in Figure 4. If the search only concentrates on some specific parts of the search space, only isolated parts of the Pareto front can be reached, which may not be good candidates for a designer's choice.

## 3.2 Multiobjective metaheuristics

This section introduces the main stream work available in the field of multiobjective optimization. We also refer the reader to a survey description of multiobjective optimization in (Collette & Siarry, 2004) and a more detailed one on evolutionary MO optimization in (Deb, 2001) for further investigations.

MO optimization techniques can be classified into the 3 following types of algorithms (Van Veldhuizen, 1999):

1/ _A priori search techniques_: A standard mono-objective search algorithms is applied to a single evaluation function that is a weighted sum of all the optimization criteria. The designer chooses the weight to assign to each evaluation function. The heuristic only provides one solution at the end of the search, reflecting the trade-off induced by the _a priori_ choices of the designer when setting the weights of the optimization criteria.

2/ _A posteriori search techniques_: A specific MO heuristic looks for the best possible approximation of the optimal Pareto front during the search. The next step is the choice of the final solution by the designer. When the estimated Pareto front is composed of too many solutions, another selection / sorting step is needed to present only some relevant solutions to the designer for a final choice.

3/ _Progressive techniques_: In these techniques, the designer directly interacts with the algorithms during the search. The search is composed of a sequence of decision making cycles (where the designer input its preferences/constraints) and search cycles. A search cycle may either use an _a priori_ or an _a posteriori_ MO search algorithm.

The Tabu based MO optimization algorithms listed in the review paper of Jones et al. (Jones et al., 2002) are _a priori_ techniques. The drawbacks of such techniques are twofold. Firstly, obtaining the trade-off targeted by the designer by choosing the weights assigned to the criteria is not always trivial. The dynamics of the optimization criteria (gradient, order of magnitude) have to be known in advance to adjust the weights properly. Depending on the complexity of the criteria, such information is not always available in real-world implementations. As a consequence, empirical trials are needed to determine the appropriate weights leading to the solution producing the desired trade-off. Secondly, for concave Pareto fronts, there may be regions of the front that are not defined by a combination of weights, and consequently certain combinations of weights represent two points on the front (Fonseca & Fleming, 1995).

Jones et al. (Jones et al., 2002) clearly demonstrated the prevalence of genetic based approaches. One reason for this is that some important theoretical algorithms have been developed in the 1980s and early 1990s for the approximation and generation of the optimal Pareto front by genetic-based methods. These include N.S.G.A., the Non-dominated Sorting Genetic Algorithm (Srinivas & Deb, 1994) and M.O.G.A., the Multi-Objective Genetic Algorithm (Fonseca & Fleming, 1998). The enhanced version of the Non-dominated Sorting Genetic Algorithm, called N.S.G.A.-II (Deb, 2002), is now recognized as one of the leading algorithms in the domain.

For these genetic based algorithms, the solutions of the population are evaluated based on a dominance metric and the non-dominated ones are stored in the estimated Pareto front. Crossover and mutation strategies are implemented in the same way as for mono-objective search to favor exploration and intensification.

The MO version of a genetic metaheuristic differs from its standard mono-objective version by the selection of the new population. This selection relies on a unique function to sort out the good solutions from the poor ones. For MO problems, the efficiency of each individual is represented by a vector of criteria. Ranking all the individuals of the population is done by a fitness function which reflects the values of all the criteria of a solution. For instance, in M.O.G.A, the solutions are ranked using an affine transformation of the Pareto-rank metric (**Definition 4)** between 0 and 1. The drawback of such a fitness function is that is does not yield a good diversity of the Pareto front.

Diversity is enforced in the fitness function of the N.S.G.A. algorithm (Srinivas & Deb, 1994) by adding a niching operator to the fitness function. This niching operator favors the selection of a solution whose evaluation belongs to an unexplored part of the current set of non-dominated solutions. This operator involves a sharing factor $\sigma_{share}$ that sets the extent of the sharing in the problem, i.e. how far any two solutions are considered to share the same fitness. The main improvement provided by N.S.G.A.-II in (Deb, 2002) is to avoid the explicit setting of $\sigma_{share}$ by providing a comparison operator that accounts for both the Pareto rank and the average distance in the function space between a solution and its neighbour non-dominated solutions.

## 4. Multiobjective Tabu (MO-Tabu)

### 4.1 Tabu heuristics for multiobjective Ooptimization

As presented in the previous section, there is a widespread interest within the engineering design community in applying multiobjective genetic algorithms to real-world problems. However, genetic algorithms can experience difficulties on highly constrained problems. Tabu search, thanks to the local search heuristic at its heart, can navigate highly constrained search spaces successfully. A multiobjective variant of Tabu search is therefore a valuable tool for engineering design. In this review, we are focusing on the *a posteriori* MO Tabu search methods where an estimation of the optimal Pareto front is targeted.

Interest for Tabu search in Multiobjective optimization has increased in the last decade (Hansen, 2000; Gandibleux & Freville, 2000; Ho et al., 2002; Armentano & Arroyo, 2004; Choobineh et al., 2006; Baykasoglu et al, 2006; Kulturel-Konak et al., 2006; Jaffrès-Runser et al., 2008; Jaeggi et al., 2008). All these techniques adapt the mono-objective Tabu search heuristic by proposing several modifications:

- Firstly, the search algorithm does not store a single estimate of the optimal solution, but a set of all the non-dominated solutions encountered during the search. At the end of the search, this set represents the estimated Pareto front. This feature is common to all the presented solutions in this section.
- Secondly, as there is not a single evaluation function to decide on the quality of the neighbour solutions, a new strategy for choosing the best neighbour at the end of one iteration has to be set.
- Thirdly, the Tabu lists are adapted to a multiobjective formulation, too. For instance, a Tabu list can store Tabu functions to avoid searching solutions along criteria already well explored (Gandibleux & Freville, 2000).
- Lastly, diversification techniques are adopted to increase the quality of the representation of the final Pareto front.

All the works presented here can be separated into two categories:

The first ones explore the space by using a single Tabu search path (Gandibleux & Freville, 2000; Ho et al., 2002; Kulturel-Konak et al., 2006; Baykasoglu et al, 2006; Jaeggi et al., 2008) while the other ones launch several Tabu searches in parallel (Hansen, 2000; Armentano & Arroyo, 2004; Choobineh et al., 2006; Jaffrès-Runser et al., 2008).

In the first case, the structure of a Tabu Search iteration is similar to an iteration of the standard mono-objective algorithm as presented in Figure 5, except for the selection of the new estimated Pareto front $\mathcal{F}_P{}^{new}$. This selection relies on a Pareto dominance test to get the non-dominated solutions from the set composed of the current Pareto front $\mathcal{F}_P$ and the set of neighbour solutions $\mathcal{V}(S_c)$.

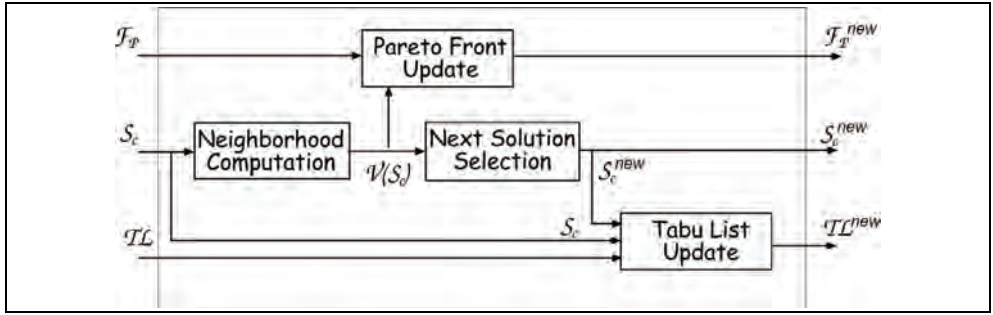

Fig. 5. Structure of an iteration for a multiobjective Tabu Search when only one search path is exploited

In the second case, there is a pool of solutions representing a search front $\mathcal{F}_c$ which is analogous to the search front defined in an evolutionary MO metaheuristic. As depicted in Figure 6, the search front is expanded through a neighbourhood search, whose solutions are evaluated and added to the estimated Pareto front to compute $\mathcal{F}_P{}^{new}$. Then a new search front $\mathcal{F}_c{}^{new}$ is selected. For these parallel strategies, the goal is to orient the parallel searches towards different parts of the Pareto front. Therefore, the selection of the new search front should orient the search to under-explored parts of the Pareto front. If $p$ search paths are considered, $p$ Tabu lists have to be updated to avoid a cyclic search for each path.

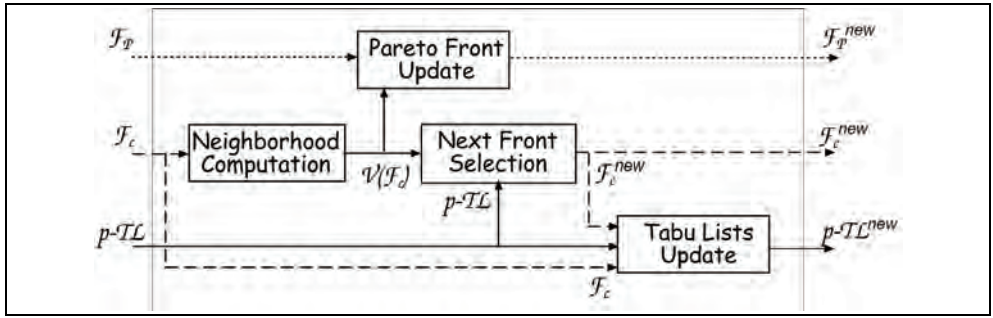

Fig. 6. Structure of an iteration for multiobjective Tabu Search when $m$ parallel Tabu search paths are exploited

**Choice of the new search solution**
The first step of a TS iteration is the creation of a set of neighbour solutions of the current search solution $S_c$. In a mono-objective problem, the new current solution $S_{new}$ is chosen as the neighbour solution that optimizes the unique optimization criterion. As a matter of fact, such a choice is not directly applicable to multiobjective problems. The whole point in a MO

Tabu search implementation is to determine what the next search step is and implicitly, which strategy will bring the search closer to unexplored parts of the Pareto front. Two kinds of strategies for the selection of $S_{new}$ have been proposed so far.

In the first strategy, a search direction in the function space is set and the solution of the neighbourhood that optimizes the objectives along that particular direction is selected. The search direction is defined by a set of weights $\{ \lambda_k,\ _{k\ \in\ [1..m]} \}$ used to optimize a weighted sum of the $m$ criteria $f_{k,\ k \in [1..m]}$ defined by:

$$f(S) = \sum_{k=1}^{m} \lambda_k f_k(S) \qquad\qquad (4)$$

For instance, Kulturel-Konak et al. (Kulturel-Konak et al, 2006) propose to randomly choose a function $f_k$ and to select the best neighbour for this particular criterion. In this case, the search direction is defined by the function for which all the weight values are equal to zero, except for the selected function which gets a weight of 1. With such a strategy, the set of search directions is equal to $m$, which necessitates more moves and therefore more function evaluations if the optimal search direction is a linear combination of the objectives. In the same way, Choobineh et al. (Choobineh et al., 2006) select a criterion as the search direction but instead of 'time-multiplexing' them, they perform a parallel Tabu search where each path optimizes one of the $m$ objectives.

Several works have focused on adjusting the weights at each solution selection to favor a rapid convergence to the Pareto front and drive the search towards unexplored parts of the front. Gandibleux and Freville (Gandibleux & Freville, 2000) favor the optimization of functions that have seen little improvement in the previous iteration. In the same fashion, Hansen (Hansen, 2000) puts more emphasis on the functions that have seen few improvements in the current search front. The difference with the implementation of Gandibleux and Freville is that Hansen also scales the weights proportionally to the number of solutions of the current search front that suffer from such a small improvement.

Armentano and Arroyo (Armentano & Arroyo, 2004) arrange the solutions of the Pareto front into clusters. Each one of their parallel Tabu search is oriented to the centroid of one of their clusters to get a good representation of the Pareto front.

The second strategy selects a solution by relying on a Pareto dominance criterion. Baykasoglu (Baykasoglu, 2006) constructs the new Pareto front $\mathcal{F}_P{}^{new}$ and then randomly selects the next solution in it.

Ho et al. (Ho et al, 2002) sort all the solutions of the search front according to their Pareto rank and apply a fitness function that favors the selection of non-dominated solutions whose neighbourhood is less dense in the function space.

**The definition of the Tabu lists**

All the heuristics use a regular Tabu list that stores for a given amount of iterations the previous solutions or movements to avoid cycling in the search. However, additional Tabu lists can also be introduced to benefit the search in the MO context.

As proposed by Gandribleux and Freville (Gandibleux & Freville, 2000), a second Tabu list is created to prevent the search along criteria that experienced consequent improvements in the past. The rationale for defining such a Tabu list is to prevent the algorithm from searching along the steepest descent gradient.

A Tabu list called the 'Intensification Memory' (IM) has been introduced in the implementation of MO Tabu by Baykasoglu et al. in 1999 (Baykasoglu et al., 1999). This memory stores non-dominated solutions that have not yet been selected as a current search solution yet. When several non-dominated solutions exist in the neighbourhood of $S_c$, the IM list is updated with the non-dominated solutions that are not chosen to become $S_{new}$. When a neighbourhood does not contain any non-dominated solution, the oldest element of the IM list is used as a new solution. Upon update, a solution is stored in the IM only if it is not dominated by any solution of the neighbourhood, by any solution of the IM list or by any solution of the estimated Pareto front. A solution of the IM list is added to the Pareto front once it has been selected as a new current solution $S_{new}$. This intensification memory list provides a tool to enforce local exploration of the solution space before storing a non-dominated solution in the estimated Pareto front.

The same memory structure has been implemented by Baykasoglu in (Baykasoglu, 2006) and by Jaeggi et al. in (Jaeggi et al.). Instead of choosing the oldest solution in the IM list, Jaeggi et al. select a random solution when intensification is needed. When the IM list is empty, Jaeggi et al. add a diversification step that randomly picks a point in the estimated Pareto front for $S_{new}$.

When $p$ parallel Tabu searches are performed, $p$ Tabu lists are maintained to avoid cycling during the search for each individual path (Hansen, 2000; Armentano & Arroyo, 2004; Choobineh et al., 2006; Jaffrès-Runser et al., 2008).

**Diversification techniques**

The search for the optimal Pareto front is challenging and diversification techniques are recommended when a single Tabu search path is considered. In this case, a rather small subpart of the solution space is explored within an iteration, resulting in an increased probability of limiting the search to a local subspace of the solution space. Hence, it is important to restart the search timely when no improvement of the Pareto front is noticed.

For instance, Kulturel-Konak et al. (Kulturel-Konak et al., 2006) restart the search with a random solution if the Pareto front remains unchanged for the last ($I_{max}/4$) iterations, with $I_{max}$ the maximal number of iterations. The Tabu list is cleared and the search resumed with the new random solution. We would like to point out that restart strategies do not necessarily induce diversification. For instance, the restart procedure proposed by Baykasoglu (Baykasoglu et al., 1999; Baykasoglu, 2006) does not introduce diversity since the new starting solution is chosen among non-dominated solutions. Therefore, only downhill moves are allowed, resulting in a local convergence.

Search diversification may also be implicitly favored through the selection of the new current solution $S_{new}$. This is the case when the search direction is adapted by changing the weights assigned to the criteria ((Gandibleux & Freville, 2000 ; Hansen, 2000). Since the weights are adapted along the search to avoid the exploration of neighbourhoods with already good performance, premature local convergence is less likely to occur.

When parallel Tabu search paths are considered (Hansen, 2000; Armentano & Arroyo, 2004; Choobineh et al., 2006; Jaffrès-Runser et al., 2008), no specific diversification procedures are required. Indeed, the number of parallel searches intrinsically increases the amount and the diversity of the solutions tested at each iteration.

Tabu search is a promising metaheuristic that efficiently tackles Multiobjective optimization. In the recent work of Jaeggi et al (Jaeggi et al., 2008), the proposed MO Tabu search implementation (called PR-MOTS) is compared to the leading genetic MO heuristic NSGA-II for a set of five standard test functions. PR-MOTS performed similarly to NSGA-II, providing a greater likelihood of improvement in the objective functions but exhibiting a higher performance variability. Moreover, the local search component and the flexibility of handling constraints for a large number of variables makes Tabu Search a particular attractive metaheuristic for multiobjective optimization.

## 4.2 MO-Tabu

A simple MO Tabu Search implementation presented in (Jaffrès-Runser et al, 2007) is described in the following to highlight the assets of Tabu algorithms for MO search. It is referred to as PMOTS, standing for 'Parallel-MultiObjective Tabu Search'.

The algorithm detailed in Figure 7 exploits $K$ parallel search paths and stores the non-dominated solutions in an estimated Pareto front $\mathcal{F}_P$. A Tabu list $TL_k$ is assigned to each search path $k \in [1,..,K]$. The duration of a Tabu list is updated at each iteration with a random size $t \in [T_{min}, \ldots, T_{max}]$.

```
1: Select the first search front Fc(0) made of K solutions;
2: Init the K Tabu lists TLk=∅, k ∈ [1,..,K];
3: Init the estimated Pareto front Fp=∅;
4: For each iteration i in [0,..,Imax] do:
5:     Init the next search front Fc(i+1)=∅;
6:     For each solution Sk of the current search front Fc(i) do:
7:          a) Compute and evaluate the neighbourhood set V(Sk);
8:          b) Select from V(Sk) the solutions with Pareto rank
             R(S) ≤ Rmax and store it as the set PR(Sk);
9:          c) Select randomly a solution of PR(Sk) and add it
             into the new search front Fc(i+1);
10:         d) Concatenate PR(Sk) with the Pareto front Fp;
11:         e) Update the Tabu list TLk;
12:    End;
13:    Remove the solutions having rank R(S)>1 from Fp;
14: End;
15: Return Fp;
```

Fig. 7. Macro-algorithm of PMOTS

In the above algorithm, the $K$ parallel search paths are represented as a search front $\mathcal{F}_c(i)$ of $K$ solutions. A new search front $\mathcal{F}_c(i+1)$ is selected in each iteration by choosing promising solutions that are not always non-dominated to avoid a premature convergence of the algorithm. Therefore for a path $k$, each new solution is selected randomly in the set of neighbour solutions $\mathcal{V}(S_k)$ of $S_k$ having a Pareto rank $R \leq R_{max}$. In this algorithm, the Pareto ranking is local to the set of neighbour solutions and does not include the current estimated Pareto set. By not including $\mathcal{F}_P$ and selecting fairly good solutions with the Pareto rank constraint, diversity is introduced within the search strategy.

The rationale behind implementing a parallel Tabu Search is the following. By properly choosing the set of initial solutions and the neighbourhood construction strategy, the optimization problem can be reduced into sub-problems being solved concurrently. Ideally, if the variable search space can be split into $K$ subsets and if the neighbourhood construction strategy of each subset is able to generate all the solutions of this particular subset, then the combination of all the $K$ parallel search paths provides a regular representation of the Pareto front.

However, each subset does not contribute equally to the Pareto front representation. A first subset might contain half of the non-dominated solutions, while another one might only have one or two Pareto optimal solutions. Therefore, if the search in the second subset is restricted to the only variables of the subset, the ratio between the cost (in terms of computation effort) and the gain (in terms of the number of solutions of the Pareto front obtained) is very poor. Therefore, to provide a more efficient search, the neighbourhood definition should not be limited to a subset of the search space, but provide some chances to move from one subset to another and thereby intensify the search in more promising subsets of the search space. The whole point in such a simple parallel optimization algorithm resides in the definition of the subsets and the neighbourhood construction process.

The first search front $F_c(0)$ is composed of $K$ randomly chosen within each subset. Depending on the MO problem considered, the neighbourhood construction process must provide most of the solutions within its own search subset and only few solutions that belong to other subsets. If a neighbour solution that belongs to another subset is promising, the search may move to that subset. This is the case when this promising solution has a Pareto rank smaller than $R_{max}$ and is selected as the new search solution.

The next section presents two implementations of PMOTS. In the first one, we present the WLP problem, and practical considerations relative to the exploitation of the estimated Pareto front are addressed here. Then, we broaden our presentation of MO-Tabu applications by discussing the problem of benchmarking the routing performance in wireless sensor networks.

## 5. Application to wireless networks optimization and evaluation

### 5.1 PMOTS for the WLP problem

PMOTS has been proposed to solve the WLP problem in (Jaffrès-Runser et al, 2007). We recall that in this problem, a solution is defined by a vector of M items $S = (s_1, \ldots, s_i, \ldots, s_M)$, each item $s_i$ representing a candidate AP location. If the candidate AP at location $i$ is selected in the solution $S$, $s_i$ stores the transmission power $p_i$ and direction of emission $d_i$ of the AP. If the AP at location $i$ is not selected in the solution, $s_i$ =0. There is a discrete set of $N_P$ possible transmission powers and a discrete set of $N_D$ possible directions of emission. At the beginning of the search, $d_i$ and $p_i$ are set to an initial value for all candidate APs.

**Neighborhood construction strategy:**

In this implementation, a neighbour of a solution $S$ is constructed by following either one of these moves:

- *Swap move*: a selected AP at position $i$ is deselected and a deselected AP at position $j$ is selected. $d_i$ and $p_i$ values remain the same as the ones already stored in the item $s_i$.
- *Addition move***:** A new AP is selected in the solution $S$,
- *Delete move:* A selected AP is deselected from the solution $S$,
- *Power change move:* Change $p$ for a selected AP to an admissible power different from the current value of $p$,
- *Direction change move*: Change $d$ for a selected AP to an admissible direction of emission different from the current value of $d$.

When computing all the neighbours according to these rules, we get the following number of neighbours as a function of the number N of currently selected APs of solution $S$:

- • N(M-N) solutions with a swap move,
- • M-N solutions with an addition move,
- • N solutions with a delete move,
- • N($N_P$ -1) solutions with transmission power change move,
- • And N($N_D$ -1) solutions with direction change move.

The solution space can be divided into subsets by gathering the solutions with the same number N of selected APs into a same subset. Consequently, a subset is defined by the number N of selected APs of its solutions and the solution space is divided into N subsets. A swap move, a transmission power change and a direction change move keep the same number of selected APs in the resulting neighbour set. The addition and deletion moves increase or reduce N, and therefore provide neighbours that belong to other subsets. The proportion of all the neighbours that belong to the same subset compared to the number of neighbours that do not belong to the same subset modifies the chances for a search path to continue to explore the old subset or to change subset.

There is an intensification of the search when the search path stays in the same subset and an exploration of the search space when the search changes subset.

For the building plan represented in Figure 1, we have M=256 candidate AP locations. There are also $N_P$=5 possible transmission power values and $N_D$=4 possible directions of emission. For this real-world implementation, how does the neighbourhood definition favor the exploration or intensification of the search depending on the number of selected neighbours N?

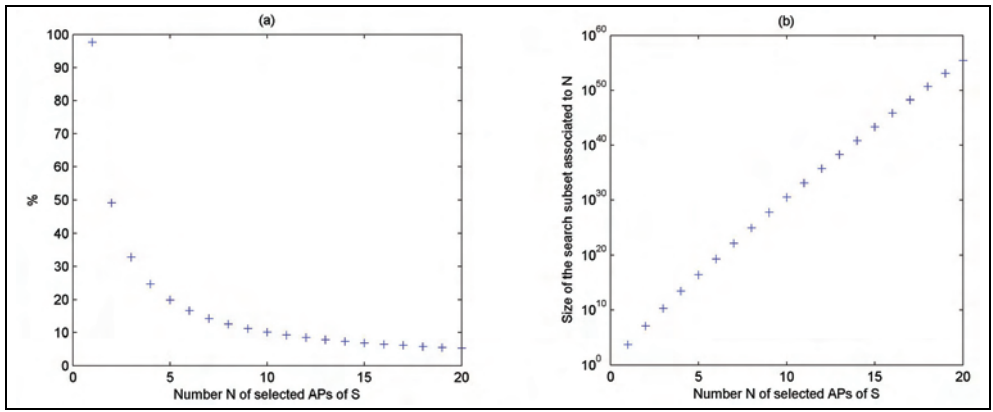

Fig. 8. Analysis of the neighbourhood $\mathcal{V}(S)$ of a solution $S$: (a) Percentage of neighbour solutions $\mathcal{V}(S)$ that belong to another subset as a function of N, the number of selected APs from the origin solutions $S$. (b) Size of the subsets having N selected APs as a function of N.

Figure 8-(a) displays the percentage of neighbours that belong to another subset depending on the number N of selected APs of the origin solution $S$. For small values of N, the percentage of neighbours with a higher number of APs is high. In this case, it is very likely that a search path changes to a subset with a higher number of selected APs. For larger values of N, there are about 5% of the neighbours that belong to other subsets, reducing the chances for a path to change subset during the search and increasing the exploration of the subset.

To summarize, for small values of N, exploration is favored while for larger values of N, intensification is favored.

Figure 8-(b) represents the size of the subsets obtained for N on a logarithmic scale varying from 1 to 20. The size of each subset increases drastically with N. Since the subsets are smaller when N is low, it makes sense to promote exploration, as a single neighbourhood construction already tests a good part of the subset. Fewer searches are here necessary in this case to explore the subset. When N increases, intensification is enforced as there are more solutions to test in a subset.

**Tabu list**

The Tabu list is a list of candidate APs. When a candidate AP is added to the Tabu list, it stores the current values of $p_i$ and $d_i$. For a swap move, the list stores the AP that is no longer selected in the solution. When the new solution results from an AP addition, the Tabu list stores a specific 'fake' AP that signifies that the addition move is now taboo. In the same way, when the solution results from the deletion of an AP, the Tabu list stores another 'fake' AP that signifies that the delete move has become taboo. When these add and delete moves are taboo, the search is forced to intensify in the current subset, which is beneficial in order to avoid too much exploration. Upon a transmission power or direction change, the candidate AP with the old value of $p_i$ or $d_i$ is stored in the list.

**About the initial solution and the number K of parallel search paths**

As there are M different subsets defined for the WLP planning problem, a first choice would be to use K=M search paths in the PMOTS algorithm. As a matter of fact, subsets composed of solutions with a high number of selected APs N can not provide very good solutions. Based on practical considerations, a planning solution that involves more than 30 APs would not perform optimally for the building described in Figure 1. For such a high number of deployed APs, the intensity of the interference in the building results in a very poor interference criterion and a reduced QoS performance. Complete coverage of the building can be achieved for N=4 APs but for a very low throughput.

Thus, we set a value of K=15 different parallel search paths with a first search path launched in the subset made of N=3 selected APs, the $k$th search path in the subset N=$k$+3 and the last search path in the subset made of N=18 selected APs.

The first search front is composed of the solutions that are the starting points for the K search paths. Each initial solution is composed of a different number of selected APs to start the search in a given subset. Consequently, the initial solution of path $k$ presents $k$+3 APs. The location of each selected AP in an initial solution is selected randomly in the set of candidate AP locations.

**Selection of the solutions of the Pareto front**

The estimated Pareto front obtained after 300 iterations is composed of 148 solutions that are presented in Figure 9. The coverage, interference mitigation and QoS criteria are optimized in this search. The target throughput for a uniform distribution of 200 users is set to 256kbits/s. For smaller instances of the WLP problem, it is shown in (Jaffrès-Runser et al, 2008) that a good approximation of the optimal Pareto front is obtained with PMOTS. In Figure 9 we have the best solutions found so far, trading-off all the planning criteria. There are 89 solutions out of 148 that have a perfect coverage and provide different trade-offs between interference and QoS. For this problem instance, PMOTS has evaluated an average of about 40000 solutions per iteration.

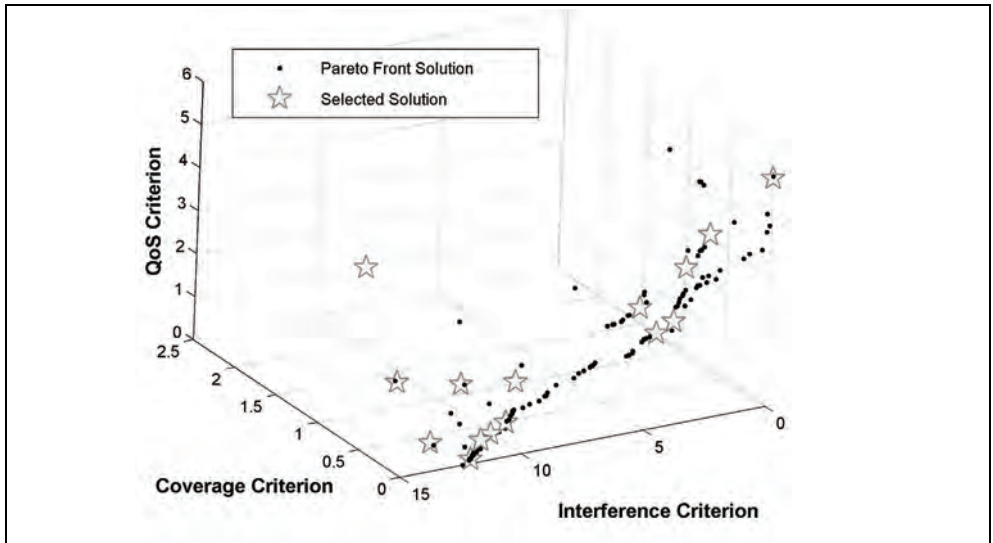

Fig.9. Representation of the estimated Pareto front for the WLP problem with M=256 candidate locations after 300 search iterations. The 15 solutions selected after the PMOTS search are represented with stars.

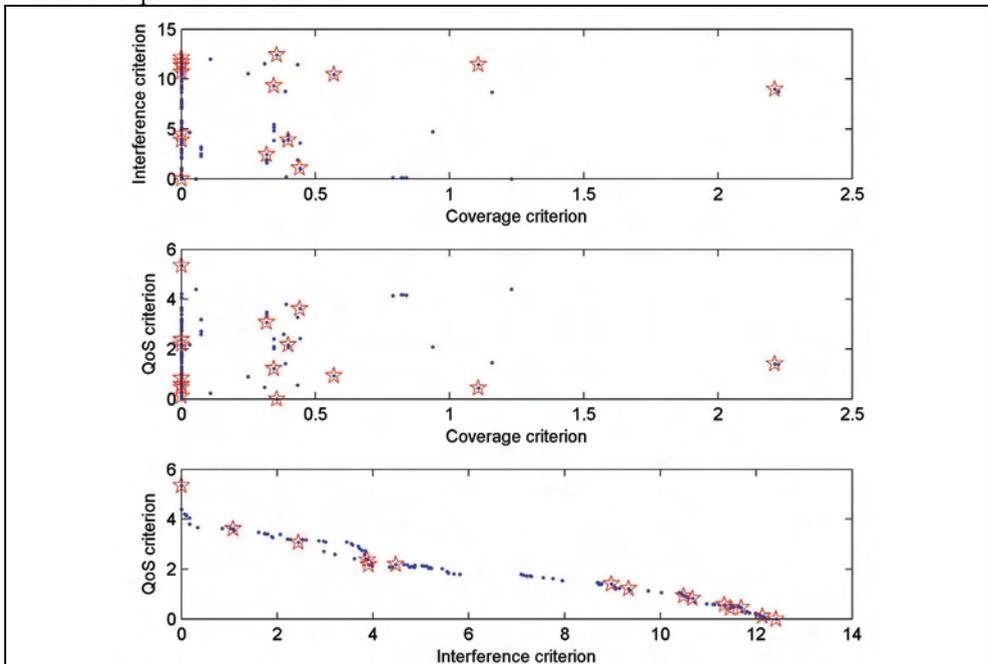

Fig.10. Projections of the estimated Pareto front for the WLP problem with M=256 candidate locations after 300 search iterations. The red stars represent the 15 solutions selected from the Pareto front.

Since the designer can not evaluate individually each one of the 148 solutions, a pre-selection of the solutions of the Pareto front is needed. An algorithm is proposed in (Jaffrès-Runser et al., 2008) which concentrates on the selection of $N_F$ solutions representing significant and different trade-offs between the objectives. The selection is based on both the dissimilarity of the criteria trade-offs obtained and the dissimilarity of the selected solutions. The $N_F$=15 selected solutions are depicted on Figure 9. The projection of the Pareto front is given in Figure 10. The 15 selected solutions are represented by little red stars.

Three network planning solutions out of the 15 selected ones are presented in Figure 11. Upon analyzing the 15 selected solutions, we concluded that the networks obtained when the number of APs is low (i.e. N<13) are very promising as they show evenly distributed APs. The solutions selected with a higher number of APs (i.e. N≥13) present an uneven distribution of APs (*cf.* Fig. 11-(c)).

Among the solutions using 19 APs, the solutions with a nonuniform distribution of nodes present a lower interference criterion than the solutions with 19 evenly distributed APs. This is due to the fact that, in the first case, strong interference is localized in a smaller surface area, thus reducing the impact on the whole building. However, as shown by the low QoS criterion, the available throughput is high even though strong interference is generated by the APs. This is due to the fact that the transmission channels are not assigned during the planning stage and therefore, the throughput estimations can not completely account for the interference distribution in the building.

Adding the channel optimization variables would increase the problem's complexity. In our case, the channel assignment is performed after the planning stage, in a separate network optimization stage. The role of the interference criterion in the planning stage is to select solutions that simplify the channel assignment step. The solution with 19 APs sees its QoS criterion rise from 0.1 to 1.0 after channel assignment as it completely accounts for interference in the throughput computation. With the increase in computation power, it is clear that including the channel assignment into the planning problem formulation will benefit the overall quality of the solutions.

However, the search space subset for N=19 APs is also bigger than the subset for lower values of N. Therefore, it might also be that the search did not found the best trade-offs that belong to this subset and that a longer search would have been beneficial for that particular path.

Introducing some interaction between the search paths should improve the performance of PMOTS. From time to time, an intensification procedure could be started where the new solution $S_{new}$ for a path $k$ with bad performance can be chosen in the neighbourhood set of promising path $j$. Every time such an intensification procedure will occur, the performance of the K paths will be evaluated in terms of the number of solutions each one of them added to the estimated Pareto front since the last intensification procedure has been performed. Upon change, the Tabu list of path $k$ would be erased as the old moves do not make sense anymore in the new search subset. By adding such an intensification procedure, we would really take advantage of the parallel processing of the PMOTS algorithm.

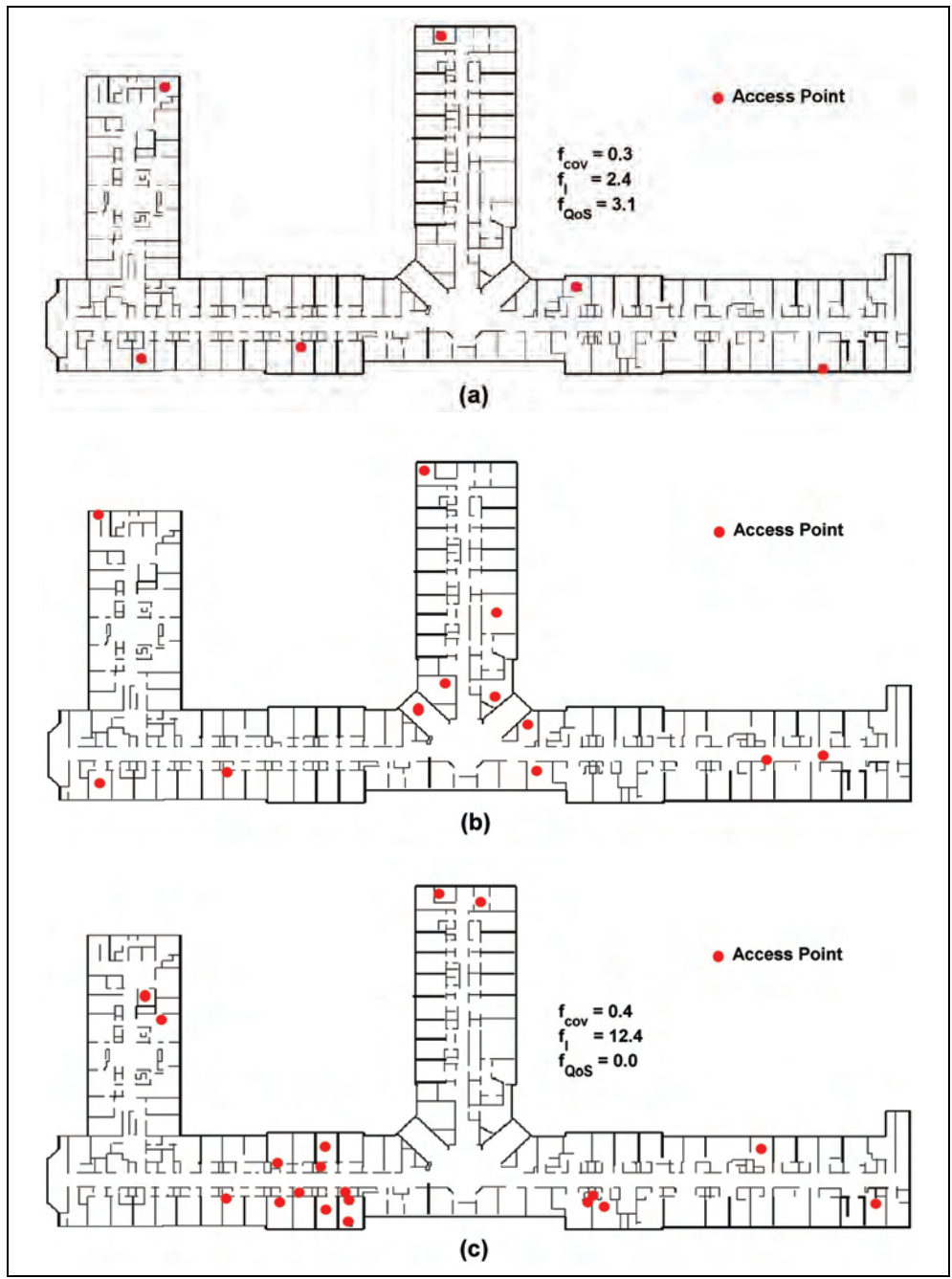

Fig. 11. Three solutions of the Pareto front: (a) Solution with 6 APs, (b) Solution with 12 APs and (c) Solution with 19 APs.

## 5.2 Application to the Evaluation of wireless sensor networks

PMOTS is adapted herein to address the problem of the evaluation of wireless sensor networks. The objective here is to quantify the achievable transmission performance of a Wireless Sensor Network (WSN). Our aim is to retrieve the Pareto optimal data forwarding patterns in the network with respect to three performance metrics: transmission robustness, transmission delay and energy consumption.

WSNs are composed of sensors equipped with a wireless transmission interface. These sensors report their sensed data to a specialized node or access point called the sink node through their wireless interface. The size of the network varies depending on the network application. Vast fields of sensor networks can be used to periodically report the activity in the field. Smaller networks can be deployed in buildings where they may sense temperature, track some target device or send alarms upon intrusion detection. Sensors are usually powered by batteries, and thus the energy consumption represents an important design criterion. Since the wireless interface consumes a significant amount of energy for listening, transmitting and receiving the data, the wireless routing protocol has a strong impact on the life duration of the network.

In this example, we introduce a model for analyzing the performance of routing in a wireless sensor network within a multiobjective framework. This model provides a tool to characterize the tradeoffs between robustness, delay and energy performance objectives depending on the network topologies and the transmitted traffic. When wireless sensors are deployed in outdoor areas, they become prone to node and transmission failures. Therefore, reliability of data transmission is a key performance metric of the network. As the data travels over the network, the average transmission delay should not affect the timeliness of the data arriving at the sink. If the transmission in the network takes too long, the data might not be accurate anymore, which results in a waste of energy.

**The multiobjective optimization model**

In this problem, a WSN consisting of N uniformly distributed sensors over an infinite plan is considered. It is assumed that the sensors are independent and randomly distributed according to a random point process of density $\rho$ over the space $\mathfrak{R}^2$. A set of $N_A$ sensors that belong to a circular area of radius R is defined as the communicating nodes of the network. These nodes can transmit, receive or forward data. The other nodes can only participate in the forwarding effort but can not be the final destination or send new data. A communication pattern is defined which is given by a set $\mathcal{S}$ made of S source nodes and a set $\mathcal{D}$ made of D destination nodes. In a WSN, we usually have D<<S as only few sink nodes exist.

The purpose of this model is to determine, given a network of density $\rho$ and a communication pattern, what kind of trade-offs arise between the transmission robustness, the transmission delay and the energy consumption depending on the routing strategy involved. Routing between source and sink can use a direct transmission (single-hop) or a multi-hop path where intermediary sensors forward the data towards the sink. In some cases, a multi-hop relaying strategy may benefit the overall energy consumption of the network as less power is needed to transmit the information on short range hops than on a single long-range hop. However, the more hops are used on a path, the longer the transmission lasts. There is a clear trade-off between the energy consumption and delay. As stated previously, transmission robustness is also of a great concern in such networks. Therefore, transmission redundancy may be also introduced by defining multiple source destination paths. Routing in a WSN is modelled as a multiobjective problem in the following. The optimization criteria are the robustness, the delay and the energy consumption of a continuous flow of data between a set of sources and a set of sink nodes.

**Optimization variables**

Each sensor node of the network can retransmit a received data packet with a given probability. This forwarding decision is modelled with a discrete variable $x_i$ representing the probability that a node forwards a message by broadcasting it. The set of P possible values for $x_i$ is given by $\{p_k$, for $k \in [1, …, P]\}$ with $p_1 = 0$ and $p_P = 1$. If P=2, $x_i$ is a binary decision variable and a sensor simply decides to broadcast a packet or not. A solution is given by the set of broadcasting probabilities:

$$\mathcal{S} = \{x_i\}_{i \in [1,..,N]} \tag{5}$$

The aim of the optimization model defined in this work is to see how the selection of the relaying nodes impacts the robustness/energy/delay tradeoffs in the WSN.

We consider that all the nodes with $x_i > 0$ constantly transmit data. For a link between two nodes $i$ and $j$, the expected interference $I_{ij}$ created at a node j can be computed by weighting the sum of the powers received at node $j$ by all the other transmitters of index $k \neq i$ by the probabilities of forwarding $x_k$:

$$\bar{I}_{ij} = \sum_{k=1}^{N} P_k . a_{kj} . x_k . \gamma_{ikj} \quad \text{for } k \neq i \tag{6}$$

In equation (6), $a_{kj}$ represents the propagation attenuation factor between node $k$ and $j$ and $\gamma_{ikj} \in [0,1]$ represents the probability that the packet transmitted by the interferer $k$ would contribute to the interference at the receiver $j$. This factor is influenced by the medium access control selection. For example, for a CDMA system using a spreading factor F, such a probability is of about $1/F$.

**Robustness** is defined as the probability that a message emitted at source $S$ successfully arrives at the destination node $D$, and is referred to as $P(R_{SD})$. Our aim is to maximize this probability. As energy and delay criteria need to be minimized, the robustness criterion is also formulated as a minimization criterion as: $f_R = 1 - P(R_{SD})$.

$P(R_{SD})$ can be defined as the probability that *the message arrives successfully in D in at most H hops, with* H→∝ . Therefore, we have:

$$P(R_{SD}) = 1 - \prod_{h=1}^{\infty} (1 - P(R_{SD}|H = h)) \tag{7}$$

where $P(R_{SD}|H=h)$ is the probability for a packet to arrive in $h$ hops at the destination. We have the probability to reach the destination in 1 hop defined as $P(R_{SD}|H=1) = p_{SD}$, the successful transmission probability on the direct link between *S* and *D*. This probability is a function of the Bit Error Rate (BER) which depends on the Signal to Interference and Noise Ratio (SINR) of the link and on the transmission technology. When more hops are accounted for, $P(R_{SD}|H=h)$ can be defined recursively as:

$$P(R_{SD}|H = h) = 1 - \prod_{j=1}^{N_S} [1 - p_{Sj} \cdot x_j \cdot P(R_{jD}|H = h - 1)] \qquad (8)$$

where $N_S$ is the number of possible first hop relay nodes of the source node S and $p_{Sj}$ is the link probability between the source node and its neighbour *j*.

Since we cannot handle an infinite number of hops in the sum of Eq.(6), we need to set a maximum number of hops allowed in the communication: $H_{max}$. Once $H_{max}$ is set, we will obtain the Pareto-optimal front for a delay-constrained network with a maximum allowed number of hops enforced. In this case, we can analyze the three-objective optimization problem knowing that the delay cannot exceed a value of $H_{max}$. The computation of the sum for Eq.(6) can also be stopped when the $P(R_{SD})$=1 (i.e. $f_R$=0). In this case, we have a two-objective problem which provides the trade-offs between delay and energy consumption when all the data is transmitted perfectly. There are solutions where $P(R_{SD})$ can never reach 1 as there are not enough paths to the destination. In this case, such a solution is dropped from the neighbourhood.

**The end-to-end transmission delay** is given by the sum of the times spend at each relay node on a multi-hop path. Since in a fixed network propagation delays are negligible, the number of relays between source and destination is a good measure of the transmission delay. The criterion $f_D$ is defined in this model as the 2nd order moment of the delay distribution among all the available paths:

$$f_D = \sum_{h=1}^{\infty} (h-1)^2 . R_h \qquad (9)$$

The quantity (*h-1*) is the delay needed by a packet to arrive in *h* hops using (*h-1*) relay nodes. $R_h$ is the probability that the packet arrived in *h* hops and did not arrive in 1, or 2… or (*h-1*) hops. For *h=1*, we have $R_h$=$P(R_{SD}|h=1)$ and for *h>1* we have:

$$R_h = P(R_{SD}|H = h) . \prod_{i=1}^{h-1} (1 - P(R_{SD}|H = i)) \qquad (10)$$

**The energy criterion** $f_E$ is defined as the average energy needed for a packet to reach its destination *D* starting from a source node *S*, whatever the path or number of hops needed. On a given path, the energy is the sum of the energy spent by all the hops that participate in the forwarding effort. We do not account for the energy needed by the source node to transmit its first packet. The average energy is the sum of the average energies needed to go from source to destination in $H = h$ with at least one path. This criterion is defined as:

$$f_E = \sum_{h=1}^{\infty} E(R_{SD}|H = h) \tag{11}$$

where $E(R_{SD}|H = h)$ is the average energy needed for a successful transmission to D in $h$ hops defined recursively as:

$$E(R_{SD}|H = h) = \sum_{j=1}^{N_s} [p_{Sj} \cdot x_j \cdot e_j + E(R_{jD}|H = h - 1)] \tag{12}$$

$E(R_{SD}|H = 1) = 0$ since the energy transmitted by the source node is not taken into account here.

**Implementation of PMOTS**

PMOTS has been adapted to address this combinatorial optimization problem. The same strategy as for the WLP problem in defining the search subsets is adopted here. Instead of gathering the solutions regarding the number of selected candidate APs, we gather them according to the number of forwarding sensors. In the same way, we can explore the subsets made of 1, 2, … or F forwarding nodes in the network. The size of each subset is given by the number of combinations of F elements out of N elements when a binary variable is considered (a sensor forwards or not).

The neighbourhood is here defined in the same way as for the WLP problem. There are swap, add and delete moves. Instead of applying these moves to the selected APs, we apply them to the forwarding nodes. Furthermore, instead of changing the direction of emission or the transmission power values, there is a probability of forwarding move where a new neighbour node sees its value of $p_k$ being changed to another possible forwarding probability.

As for the WLP problem, solutions where numerous nodes forward at the same time are not profitable as they generate a lot of interference, reducing drastically the probability of correct reception while increasing inefficiently the energy consumption. Therefore, we also favor the search for solutions with a low number of forwarding nodes by choosing accordingly the starting solutions of the search front.

In the simple case where we only have one source-destination communication, it is clear that only a few relays are useful. When the number of concurrent communications increases, more sensors have to contribute to the forwarding effort. Therefore, for every problem instance, a good choice of the starting solutions is beneficial for the search.

In this chapter, we simply address the problem of a single source-destination transmission. A node density of $\rho = 0.7$ is considered for a network consisting of N=334 nodes. For this particular problem instance, we set the maximum number of hops to $H_{max}=4$ and minimize robustness, delay and energy. There are 4 parallel search paths and the number of forwarding nodes is limited to 4. The initial front is composed of solutions with F=1, F=2, F=3 and F=4 solutions. The estimated Pareto front obtained after 1000 iterations is presented in Figure 12.

This Pareto front shows the trade-offs between robustness, delay and energy. A zero delay and energy is obtained for a reliability of 0.245. This particular solution reflects the direct source-destination communication where no other node is forwarding. For the solutions with the highest values of energy, four relays actively participate in the broadcasting effort and achieve the best possible robustness. For all the other solutions, the number of broadcasting relays varies between 1 and 4 depending on the link quality.

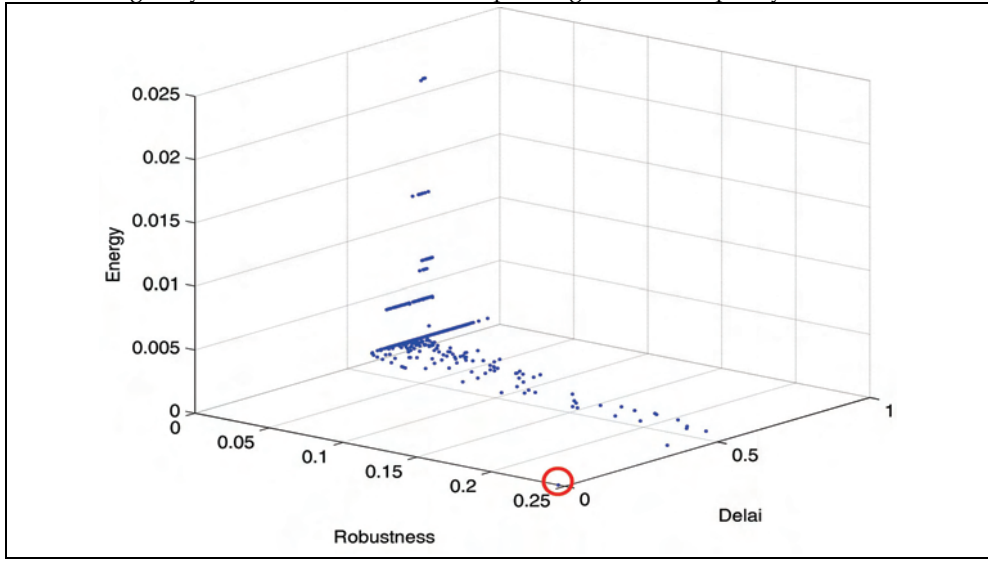

Fig.12. Estimated Pareto front for the single communication performance estimation problem

This representation of the Pareto front for the single communication problem is obtained after only about 635000 function evaluations. However, if more source destination pairs are considered in the network, solutions with a larger number of forwarding nodes are needed. In this case, the search can not simply target small parts of the solution space and a probably longer search would be necessary to investigate subsets with a higher number of forwarding nodes. As a consequence, we infer that the intensification strategy proposed at the end of section 5.1 will be very beneficial for the resolution of this problem as a more accurate estimate of the Pareto front is needed here.

## 6. Conclusion

As presented in this chapter, Tabu Search is a promising metaheuristic when addressing multiobjective optimization problems. It is particularly suited to handle problems with numerous combinatorial or continuous variables. The local search at its heart also makes it an interesting technique for highly constrained optimization problems. Most of the heuristics dealing with MO Tabu search use a single search path.

PMOTS, the algorithm used as an example in this work, relies on a simple parallel search which uses several paths and a specific neighbourhood construction strategy in order to spread the search paths in all the interesting parts of the solution space. The benefits of such an approach are highlighted for two particular problems arising in wireless systems. In the first one, a real world WLAN network has to be planned to meet several performance guarantees. In the second one, a more theoretical use of the PMOTS algorithm is presented. In this particular case, the search really needs to provide the best possible estimate of the Pareto front for the results to be meaningful from the theoretical point of view. This contrasts with the WLAN planning problem where the search is asked to provide good solutions (but not optimal ones) in a limited search time. PMOTS performs very well for the WLP problem. However; the search time for convergence to the optimal Pareto front still needs some improvements. Therefore, an intensification stage that disregards bad performing paths and intensifies the search for promising ones might be beneficial to increase the convergence speed of the parallel approach.

From a more general point of view, multiobjective Tabu search is still a promising research area. The Tabu list at its core provides a mean to efficiently explore a good portion of the search space. Clever uses of Tabu lists that store information about the past solution evaluations have been proposed which improve the representation of the estimated Pareto front. Since Tabu search is easy to implement and performs nicely on a large set of single objective optimization problems, it makes it a good candidate for the fast implementation of a range of multiobjective real-world problems.

## 7. Acknowledgement

The work presented in this chapter has been supported by the Marie Curie program from the European Community's Sixth Framework Program. This chapter only reflects the Author's views and the European Community is not liable for any use that may be made of the information contained herein.